\documentclass[superscriptaddress,showpacs,twocolumn,amssymb,aps]{revtex4}

\usepackage{graphicx,dcolumn,bm,amsmath,color,hyperref}

\newcommand{\eq}[1]{Eq.~(\ref{#1})}

\newcommand{\eea}{ \end{eqnarray} }
\newcommand{\bea}{ \begin{eqnarray} }

\newcommand{\eeq}{ \end{equation} }
\newcommand{\beq}{ \begin{equation} }

\newcommand{\bhua}{ \hat{\bf u} _{\alpha} }
\newcommand{\bhub}{ \hat{\bf u} _{\beta} }

\newcommand{\bhu}{ \hat{\bf u} }

\newcommand{\bra}{ {\bf r}_{\alpha}  }
\newcommand{\brb}{ {\bf r}_{\beta}  }

\newcommand{\bs}{{\bf s }}

\begin{document}


\title{Controlling active self-assembly through broken particle symmetries}

\author{H. H. Wensink}
\email{wensink@lps.u-psud.fr}
\affiliation{Laboratoire de Physique des Solides, Universit\'{e} Paris-Sud \& CNRS, B\^{a}timent 510, 91405 Orsay Cedex, France}

\author{V. Kantsler}
\affiliation{DAMTP, Centre for Mathematical Sciences, University of Cambridge, Wilberforce Road, Cambridge CB3 0WA, UK}

\author{R. E. Goldstein}
\affiliation{DAMTP, Centre for Mathematical Sciences, University of Cambridge, Wilberforce Road, Cambridge CB3 0WA, UK}

\author{J. Dunkel}
\affiliation{DAMTP, Centre for Mathematical Sciences, University of Cambridge, Wilberforce Road, Cambridge CB3 0WA, UK}

\date{\today}

\begin{abstract}
Many structural properties of conventional passive materials are known to arise from the symmetries of their microscopic constituents. By contrast,  it is largely unclear how the interplay between cell shape and self-propulsion controls the meso- and macroscale  behavior of active matter. Here, we analyze large-scale simulations of homo- and heterogeneous  self-propelled particle systems to identify  generic effects of broken particle symmetry on collective motion. We find that even small violations of fore-aft symmetry lead to fundamentally different collective behaviors, which may facilitate demixing of differently shaped species as well as the spontaneous formation of stable micro-rotors. These results suggest that variation of particle shape yields   robust physical mechanisms to control self-assembly of active matter, with possibly profound implications for biology and materials design. 
\end{abstract}

\maketitle

\begin{figure}[b]
\begin{center}
\includegraphics[clip=,width= 0.87\columnwidth]{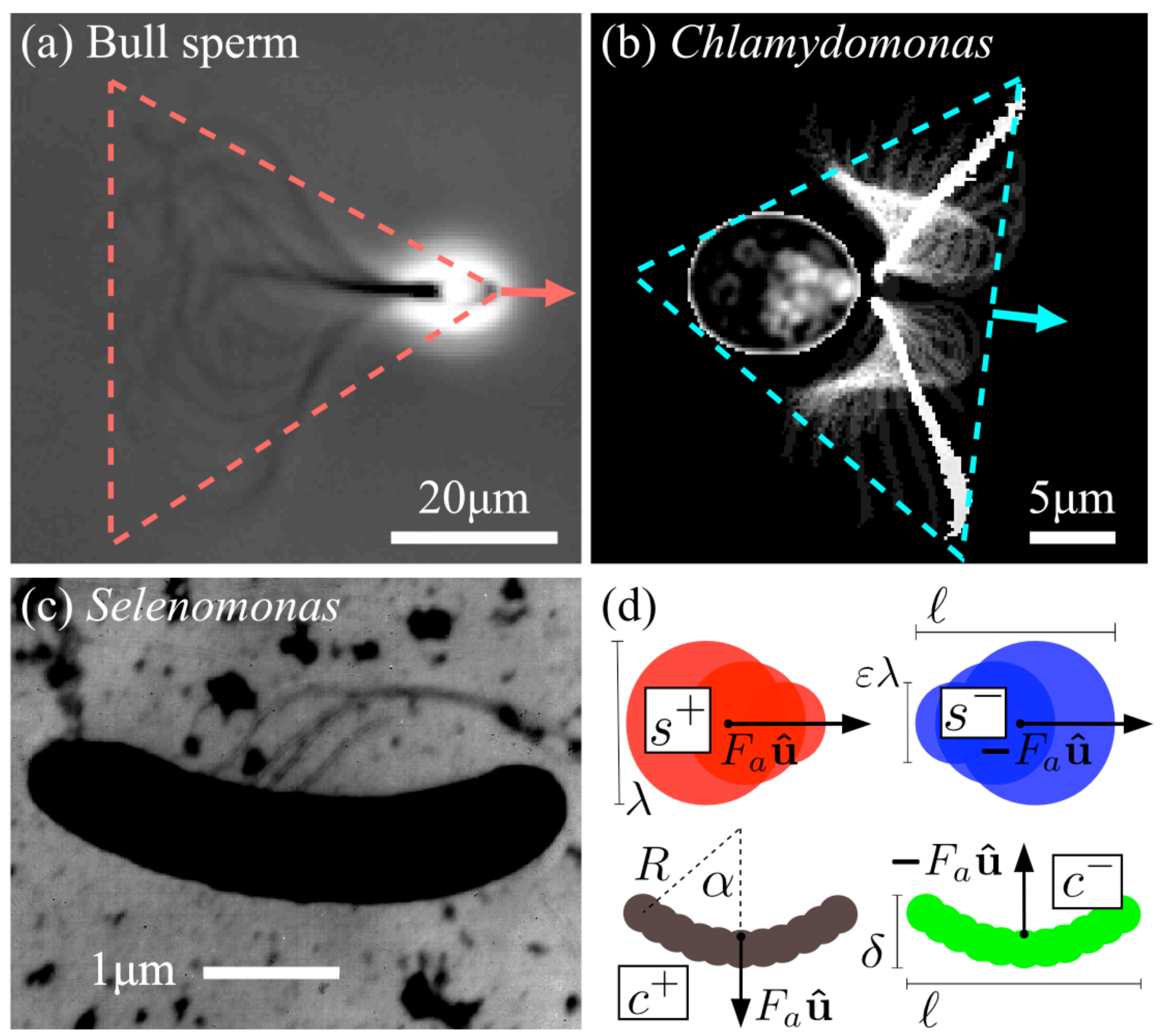} 
\caption{ \label{fig01} 
(color online)  Actual and effective dynamical shapes of microorganisms, and their simplified representation in the SPP model.  (a)~Superimposed phase-contrast micrographs (Zeiss Axiovert, 40$\times$, NA0.6) of swimming bull sperm. On time-scales larger than the beat period $\sim0.1$s, the cell mimics a forward-swimming cone. (b)~A \textit{Chlamydomonas} alga (63$\times$, NA1.3), confined to quasi-2D motion, resembles a backward-swimming triangle.  (c)~Non-convex crescent shaped \textit{Selenomonas bovis} bacterium  with flagella; reprinted with kind permission from Ref.~\cite{2009Selenomonas_IJSEM}.  (d)~The SPP model approximates different shapes by combinations of rigidly linked spheres. 
}
\end{center}
\end{figure}

Physical and chemical properties of conventional materials depend critically on the symmetries of their microscopic constituents~\cite{1929Pauling,2012Glotzer_Science,2012Graaf_Science,2012Dogic_Nature, 2012Bausch_Nature}. The perhaps best-known examples are carbon allotropes~\cite{2010Hirsch_NatMat}, such as diamond or graphite, which exhibit vastly different  elastic and optical characteristics reflecting the tetrahedral or planar structure of their elementary building blocks. Whilst the relation between microscopic symmetries and macroscopic properties of passive materials has been intensely studied  both experimentally~\cite{2003Mao_Science, 2012Dogic_Nature,2012Wang_Science} and theoretically~\cite{1929Pauling,2012Glotzer_Science,2012Graaf_Science,2012Cegeo_EPL}, comparatively  little is known about how constituent shape affects structure formation and collective motion in active matter~\cite{2010Ramaswamy,2012Menzel_EPL,2012Wensink_PNAS,LaugaGoldstein2012,2013Lauga_PRL, 2013Dunkel_PRL}. Rapid progress in the fabrication of colloids~\cite{2007ColloidShape_Review} and Janus particles~\cite{2013Janus_Review} with broken spherical~\cite{2013Shelley_Rods_PRL} and cylindrical~\cite{2013Loewen_PRL}  symmetries suggests the need for systematic theoretical studies to help  identify generic principles for the design and self-assembly of novel active matter states. Additional biophysical motivation stems from the still limited understanding about the role of cell-shape~\cite{2006Young,2009Capeen_EMBO} in the evolution of multicellular behavior~\cite{2012Wensink_PNAS,2013Beer_JBac}. Although the relative importance of chemical, hydrodynamic and steric interactions for collective microbial motion is still under debate~\cite{2011DrescherEtAl,2013Aranson_Physics},  recent experiments~\cite{2013Kantsler_PNAS} have shown that the different surface-scattering laws of sperm and \textit{Chlamydomonas} cells arise from their effective dynamical shapes, that is the volumes swept by their cilia during swimming [Fig.~\ref{fig01}(a,b)]: Sperms can be viewed as forward-swimming cones that tend to accumulate at surfaces since, after an aligning collision, their velocity vector points into the boundary. By contrast, biflagellate \textit{Chlamydomonas} algae resemble backward-swimming cones, departing from lateral boundaries at a narrowly distributed angle set by the ciliary beat~\cite{2013Kantsler_PNAS}. These observations raise the question whether similar shape-induced steric effects suffice to explain various types of collective microbial dynamics.

\par
In this Letter, we show that even small violations of fore-aft symmetry may lead to fundamentally different modes of collective motion in active systems. By analyzing large-scale simulations of two-dimensional (2D) self-propelled particle (SPP) systems, we find that purely shape-induced interactions lead to  front-like cooperative motion of sperm-type swimmers, whereas alga-like swimmers tend to cluster in structures that resemble multicellular colonies. These qualitatively different behaviors facilitate spontaneous demixing of inhomogeneous suspensions, suggesting that the combination of particle shape and self-propulsion might have been a relevant  evolutionary factor and also offering robust tuning mechanisms for the self-assembly of active materials.  To illustrate the latter fact, we will demonstrate that non-convex  SPPs [Fig.~\ref{fig01}(c,d)] can self-assemble into active rotors. 

\paragraph*{Model.}
We simulate  $N$ SPPs in 2D, each driven by a constant self-propulsion force $F_{a}$ of  fixed direction in the body frame.  The 2D case is practically relevant as colloids and microorganisms often accumulate at surfaces and interfaces~\cite{2009So,DiLeonardo2010,2013Kantsler_PNAS}. Focussing on two important classes of shapes, we compare convex sperm-type SPPs ($s^+$) with \lq antisperms\rq~($s^{-}$),  and  non-convex crescent-shaped $c^+$-SPPs with \lq anticrescents\rq~($c^-$), as defined in Fig.~\ref{fig01}(d). Particles are assumed to move in the overdamped
low-Reynolds number regime~\cite{1977Pu}, interacting with each other only by  steric repulsion. Interparticle forces and torques are calculated by discretizing each SPP into $i=1,\ldots,n$ equidistant spherical segments with effective diameter~$\lambda^i$. Defining~$\lambda^{ij}_{\alpha \beta}= (\lambda^i_\alpha+\lambda^j_\beta)/2$,  the total pair potential  $U_{\alpha \beta} =  n^{-2} \sum_{i,j=1}^{n} u( r^{ij}_{\alpha \beta}/\lambda^{ij}_{\alpha \beta})$ of two SPPs $\alpha$ and $\beta$ depends on their orientation unit vectors $\{ \bhua, \bhub \}$ and  center-of-mass distance $\Delta {\bf r}_{\alpha \beta}=\bra - \brb $ through the segment distance $r^{ij}_{\alpha \beta} = |\Delta {\bf r}_{\alpha \beta} + \bs^{i}_{\alpha}  - \bs^{j}_{\beta} |$, where the vectors $\bs^{i}_{\alpha}$ denote the position of segment $i$ relative to the mass center $\bra$ with respect to the body frame.  Throughout, we adopt a repulsive short-range potential $u(x)=u_{0}\exp(-x )/x^{2} $ with amplitude $u_{0} >0$ and consider minimal deterministic equations of motion for the positions $ \bra(t)$
and orientations $\bhu_\alpha(t) =  \{ \sin \varphi_\alpha(t) , \cos\varphi_\alpha(t) \} $ by balancing active and steric forces and torques,
\begin{eqnarray}
{\bf f }_{ T} \cdot \partial_{t} \bra =  -\nabla_{\bra }
 U +  F_{a} \bhua , 
 \quad
{\bf f}_{R} \cdot \partial_{t} \varphi_{\alpha} =
-\nabla_{\varphi_{\alpha}} U ,
\label{eom}
\end{eqnarray}
where  $U=(1/2)\sum_{\alpha, \beta (\alpha \neq
  \beta)} U_{\alpha \beta}$ is the total potential energy. The one-particle translational and rotational friction tensors ${\bf f}_{T} $ and ${\bf
  f}_{R}$ can be decomposed into  parallel, perpendicular
and rotational contributions which depend solely on the
aspect ratio $a$ for which we specify below effective values depending on the SPP shape
\cite{tirado}.   Eq.~\eqref{eom} neglects thermal or intrinsic Brownian noise~\cite{2011DrescherEtAl}, which is acceptable at intermediate-to-high concentrations when particle collision dominate the dynamics~\cite{2012Wensink_PNAS}.
\par
We integrated~\eq{eom} numerically using a square simulation box (area $A$) with periodic boundary conditions and random initial conditions. The distance between neighboring SPP segments  was kept small to avoid crossing of  SPPs [Fig.~\ref{fig01}(d)].  To reduce the number of parameters in simulations, Eq.~\eqref{eom} was rewritten in terms of the dimensionless time $\tau = t v_{0}/\lambda$, adopting  the largest segment diameter  $\lambda$ as length unit and the self-propulsion speed $v_{0}=F_{a}/||{\bf f }_{T}||$  of a noninteracting SPP as velocity unit.  The rescaled potential amplitude $\tilde{u}_{0}=u_{0}/(F_{a}\lambda)$  has  little or no effect on the collective behavior as long as $\tilde{u}_{0} > 10$ and the SPP dynamics is governed by the volume fraction $\phi = N\sigma/A $, with $\sigma$ the area per particle, and the intrinsic shape parameters of the SPPs (aspect ratio, polarity, curvature, etc.).

\begin{figure*}[!]
\begin{center}
\includegraphics[clip=,width= 2.0\columnwidth]{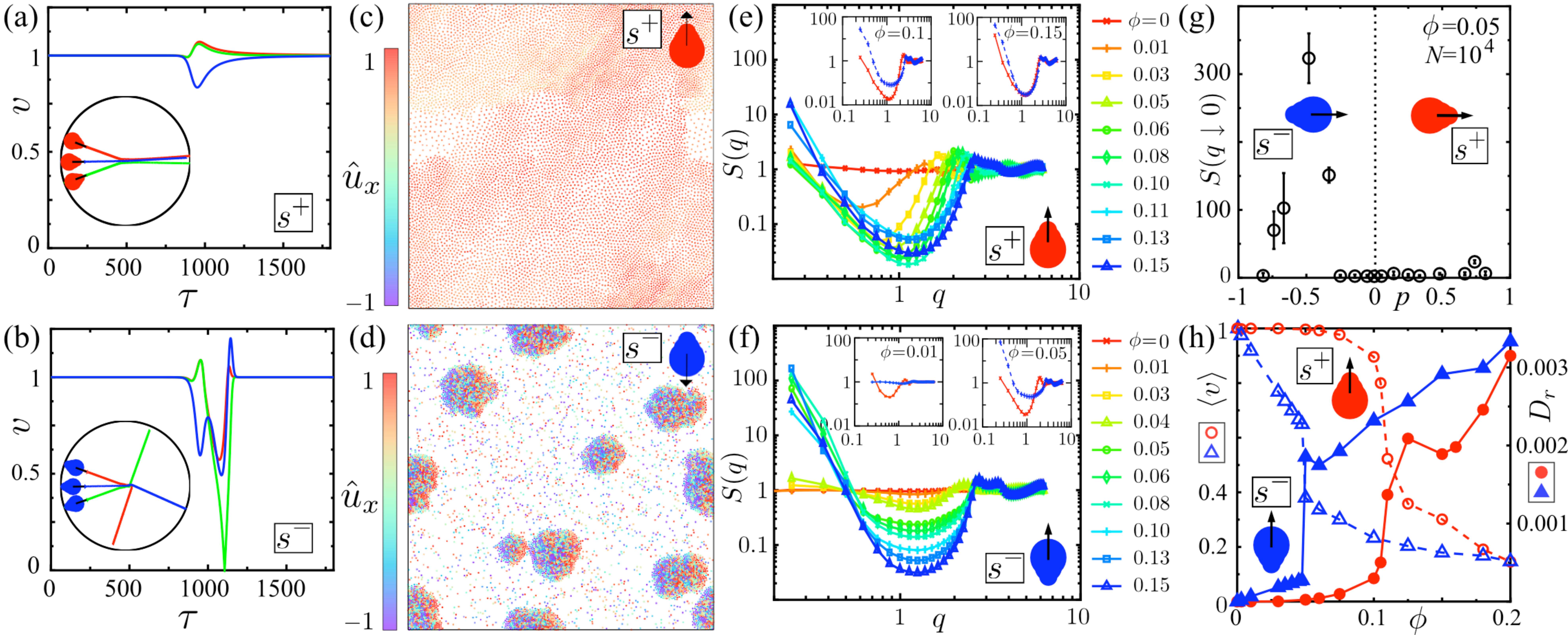} 
\caption{\label{fig02} 
(color online) Simulation results for uniform suspensions of convex particles, see also \href{https://dl.dropboxusercontent.com/u/101166552/MOVIE_01.mp4}{Supplemental Movies 1} and \href{https://dl.dropboxusercontent.com/u/101166552/MOVIE_02.mp4}{2}. 
(a,b)~ Time-dependent speed $v$ in units of $v_0$ and trajectories (insets) of 3 colliding sperm-type $s^+$-SPPs and antisperm $s^-$-SPPs.
(c,d)~$s^+$-SPPs form aligned fronts, whereas $s^-$-SPPs exhibit clustering. Color encodes the horizontal orientation components $\hat{u}_{x}$  ($N=10^{4}$, $|p| =0.33$, $\phi = 0.05$).  (e,f)~Static structure factor $S(q)$ at different volume filling fractions $\phi$.  Insets: Comparison of $S(q)$ for $s^+$-SPPs (red) and $s^-$-SPPs (blue) at two different  filling fractions $\phi$. (g) The peak of $S(q\to 0)$ at $p\approx-0.5$ indicates an optimal polarity for cluster formation. (h)~Average velocity $\langle v \rangle $ for SPPs with $p= \pm 0.33$ (dashed/open symbol) and rotational diffusion coefficient $D_{r}$ of a tagged SPP (solid/filled symbols). }
\end{center}
\end{figure*}

\paragraph*{Convex SPPs.}
We first consider convex sperm-type $s^+$-SPPs and antisperm $s^-$-SPPs  [Fig.~\ref{fig01}(d)], composed of $n=3$ equidistant spherical segments and representing prototypical polar swimmers with broken fore-aft symmetry,  $U_{\alpha \beta}(\bhua , \bhub ) \neq U_{\alpha \beta}(\bhua, -\bhub)$.  The geometric polarity  $p  = (\lambda^{a} - \lambda^{f})/(\lambda^{a} + \lambda^{f}) $ is quantified in terms of the dimensions $\lambda^{f/a}$ of the fore/aft segments,  so that $p>0$ for $s^+$-SPPs, $p<0$ for $s^-$-SPPs, and $p=0$ for apolar rodlike SPPs. The effective aspect ratio is defined by $a=\ell/\lambda = 1+\varepsilon/2$, with $\varepsilon \in [0.1,1] $ and $p \in [-0.8,0.8]$ in simulations.

\par
The broken fore-aft symmetry results in fundamentally different collective behaviors of $s^+$-SPPs and $s^-$-SPPs, caused by their qualitatively different steric collision laws [Fig.~\ref{fig02} and Supplemental Movies 1 and 2]. Sperm-type $s^+$-SPPs tend to align and experience only small speed changes during collisions, whereas $s^-$-antisperms  scatter broadly and experience a strong reduction of their speeds during the collision process [Fig.~\ref{fig02}(a,b)].  On the mesoscopic level, these two collision scenarios translate into distinctly different patterns [Fig.~\ref{fig02}(c,d)].   $s^+$-SPPs  form aligned large-scale swarms that move cooperatively along  a spontaneously chosen common axis, whereas   $s^-$-antisperms  tend to form droplets that  nucleate slowly from an initially homogeneous suspension.   Whilst this droplet formation 
may appear visually similar to chemotactic aggregation~\cite{2008Armitage,2008Pawel}, the underlying mechanism is very different as $s^-$-interactions are purely repulsive.   

\par
The clustering instability can be quantified in terms of the static structure factor, defined as $S(q) = \frac{1}{N} \langle \rho_{ {\bf q}}(\tau)  \rho_{ -{\bf q}}(\tau) \rangle $, which is directly related to the number fluctuations in the limit of vanishing wavevector $q$ via $S(q \downarrow 0 ) = (\langle N^{2} \rangle  - \langle N \rangle ^{2})/\langle N \rangle $~[Fig.~\ref{fig02}(e,f)].  For  alga-like $s^-$-SPPs, a discontinuity in $S(q \downarrow 0)$ marks the onset of the clustering instability  at a volume filling fraction of $\phi \sim 0.05$ ~[Fig.~\ref{fig02}(f)].  This transition is significantly weaker  for  the sperm-type $s^+$-SPPs~[Fig.~\ref{fig02}(e)].  In particular, plotting $S(q \downarrow 0 )$ as a function of polarity $p$ reveal that $p\approx -0.5$ is optimal for the self-assembly of clusters~[Fig.~\ref{fig02}(g)], corroborating that  for convex polar SPPs the propensity to cluster depends crucially on the self-propulsion direction relative to the broken fore-aft symmetry. The differences in the collective behavior of   $s^+$-SPPs and $s^-$-SPPs become  most prominent at intermediate packing fractions~$\phi\sim$ 5\% to 10\%, a regime that can be achieved in suspensions of swimming microbes~\cite{2012Sokolov} and active colloids~\cite{2013Buttinoni}. At very low or very high values of $\phi$,  these differences vanish since the SPPs become effectively non-interacting (low $\phi$) or  too strongly hindered due to packing effects (large $\phi$).  

\par
To characterize in more detail the effects of shape on collective motions in homogeneous suspensions,  we measured the mean speed $\langle v\rangle$ and the effective rotational diffusion coefficient $D_{r} = \lim_{\tau  \rightarrow \infty}  \langle (\Delta \varphi (\tau) )^{2} \rangle /(2\tau)$ from the mean-squared  displacement (MSD) of the orientation angle~$\varphi$. 
When plotted as a function of the filling fraction $\phi$, both $\langle v\rangle$  and $D_r$ exhibit steep jumps for $s^-$-antisperms while varying much more smoothly for $s^+$-SPPs [Fig.~\ref{fig02}(h)]. These jumps signal the onset of a non-equilibrium phase separation for $s^-$-SPPs. A similar phenomenon was reported recently by Farell~\textit{et al.}~\cite{farrell2012} for a generalized Viscek model, suggesting that this may be a generic feature of active systems with  density-dependent mobility~\cite{2013Buttinoni}.

\begin{figure}[h!]
\begin{center}
\includegraphics[clip=,width= 0.93\columnwidth]{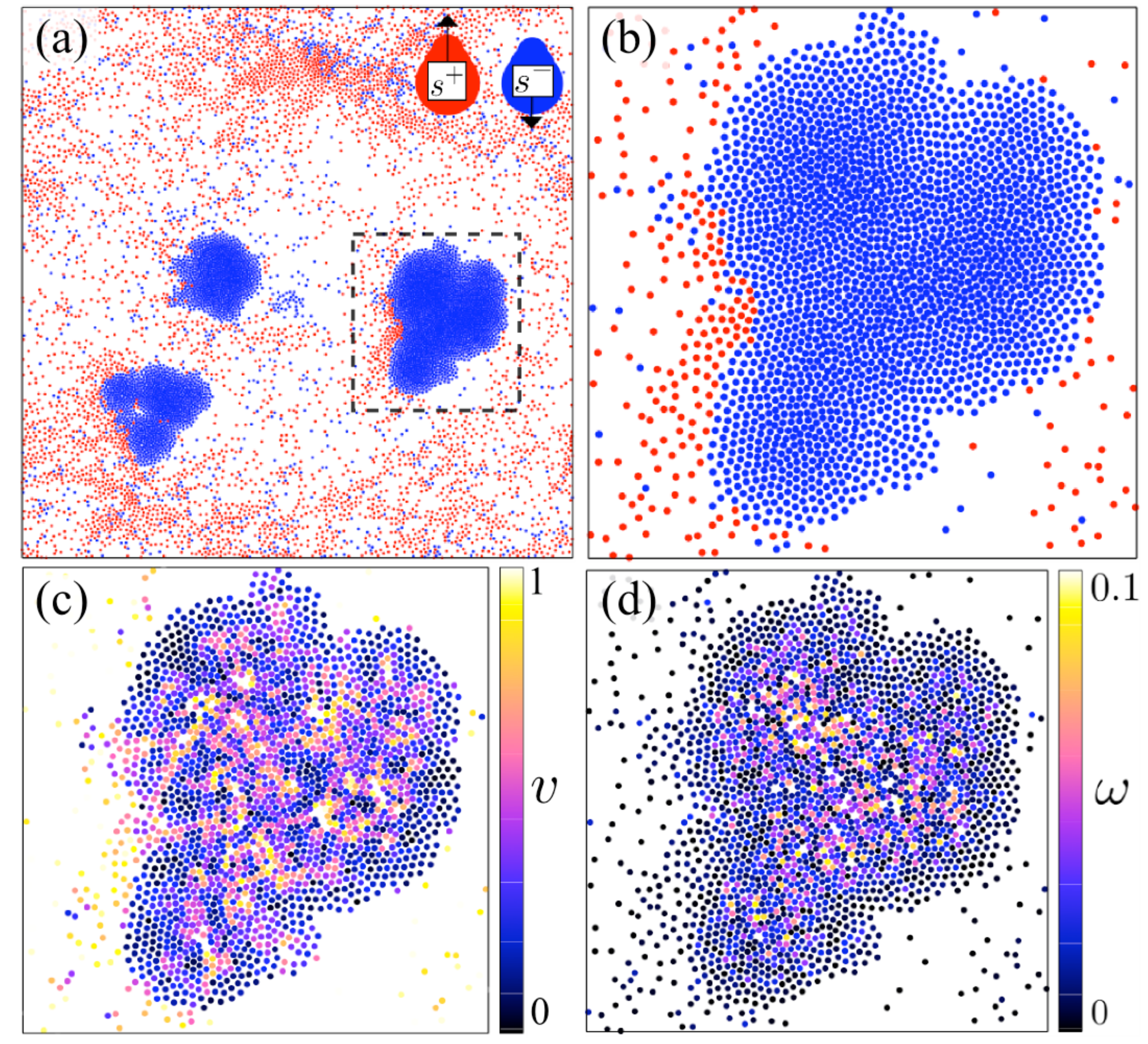}
\caption{ \label{fig03} 
(color online) Demixing of $s^+$-SPPs and $s^-$-SPPs in an equimolar  binary suspension, see also \href{https://dl.dropboxusercontent.com/u/101166552/MOVIE_03.mp4}{Supplemental Movies~3} and \href{https://dl.dropboxusercontent.com/u/101166552/MOVIE_04.mp4}{4}.
(a)~Snapshots of the center-of-mass positions for $N=10^{4}$ SPPs with $|p|=0.67$ and $\phi = 0.05$.  (b-d)~Region enclosed by dashed box in~(a).  Color in (c,d) encodes  the translational speed $v$ and rotational velocity  $\omega = | \partial \varphi / \partial \tau | $ of each SPP. }
\end{center}
\end{figure}

\paragraph*{Demixing.}
The qualitatively different collective behaviors of  sperm-like $s^+$-SPPs and Chlamydomonas-like $s^-$-SPPs suggest a simple shape-induced mechanism for demixing in active suspensions, which could be of relevance for the segregation of species in microbiological systems. To test this idea, we simulated binary mixtures with equal numbers of $s^+$-SPPs  and $s^-$-SPPs. These simulations showed that such systems do indeed segregate into dense droplets  of $s^-$-SPPs that are almost completely devoid of $s^+$-particles [Fig.~\ref{fig03}(a,b)].  The particle motions within the colony-like  droplets exhibit clear signatures of dynamic heterogeneity, as $s^-$-SPPs tend to move faster in the core of a droplet [Fig.~\ref{fig03}(c,d)].   In contrast to the surrounding $s^+$-swarms,  the total net velocity of each $s^-$-colony is very small so that  they are virtually immobile.  Moreover, it could be observed that $s^+$-SPPs tend to accumulate at the outer regions of the droplets [Fig.~\ref{fig03}(b)], reminiscent of bacterial cells accumulating on the surfaces of algal colonies. In addition to their potential biological implications, these findings further demonstrate that mixtures of suitably shaped convex SPPs can provide a basis for the targeted self-assembly of active colloids in layers or shells. 

\paragraph*{Non-convex SPPs.}
To explore the potential of another important classes of particle shapes for the self-assembly of active matter, we complement the above considerations by discussing the case of non-convex particles, using crescent-shaped $c^+$-SPPs and $c^-$-anticrescents [Fig.~\ref{fig01}(d)] as representative examples. Non-convex self-propelled colloids were recently realized in experiments by K\"ummel \textit{et al.}~\cite{2013Loewen_PRL}, and non-convex shapes can also be found in various bacteria~\cite{2009Selenomonas_IJSEM,2009Capeen_EMBO}, see Fig.~\ref{fig01}(d),  but their collective behavior  has not yet been systematically investigated.  In our simulations, we implemented crescent-shaped SPPs composed of $n=\lfloor 2s/\lambda \rceil $ overlapping spherical segments (diameter $\lambda$), equidistantly spaced on a circular arc of fixed length $s=2\alpha R$. We quantify the degree of non-convexity through the dimensionless curvature parameter   $\kappa = \lambda/R=  2 \lambda \alpha/s  $, defined such that straight rods are recovered in the limit $\alpha \to 0$ at constant arclength~$s$.  
The effective aspect ratio $a =\ell/\delta$ of the $c^\pm $-SPPs is determined by the dimensions $\ell = \lambda + s \sin \alpha /\alpha $ and $\delta = \lambda +  s ( 1- \cos \alpha )/2\alpha$.
\begin{figure}
\begin{center}
\includegraphics[clip=,width= 0.93 \columnwidth]{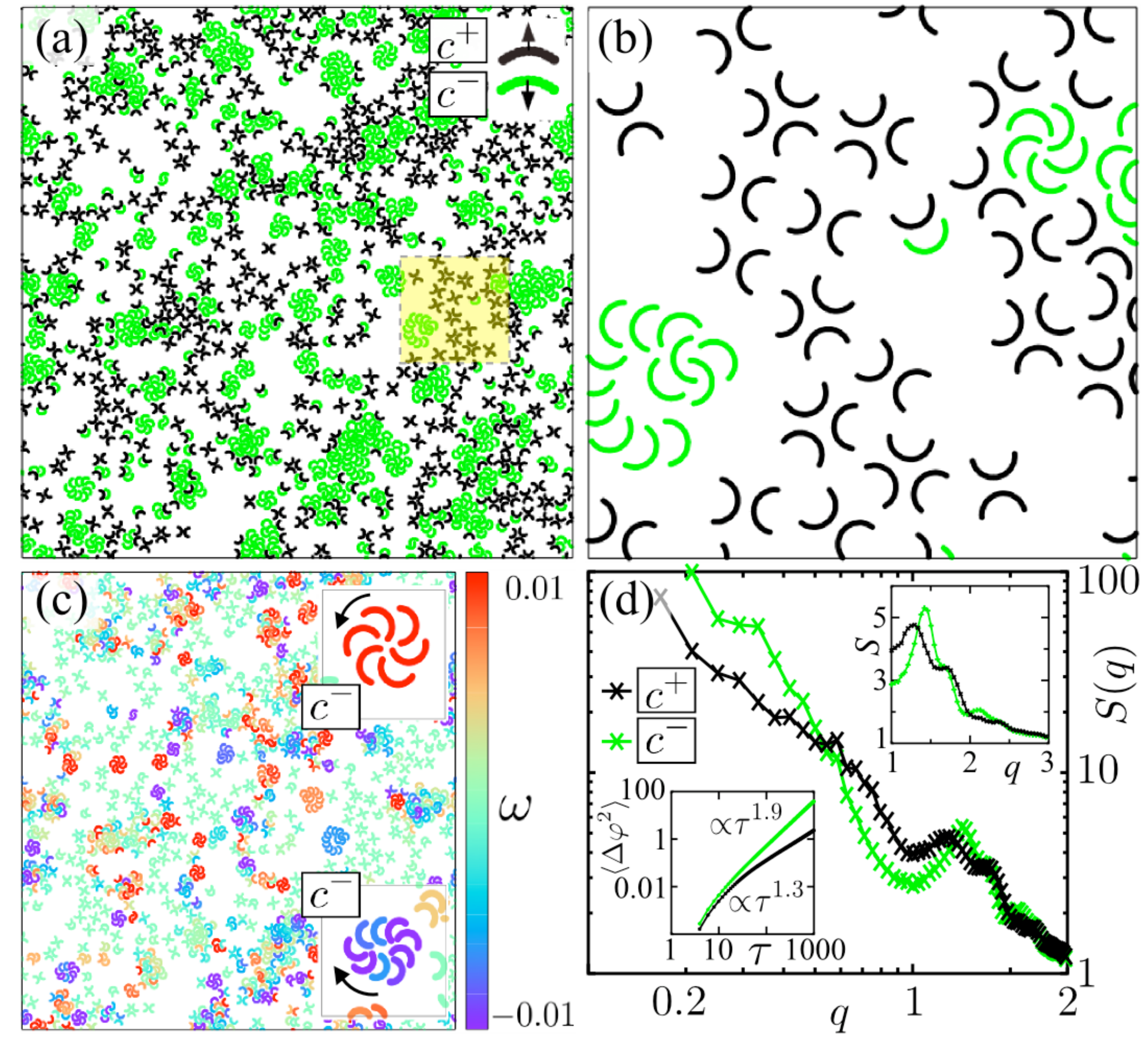}
\caption{ \label{fig04} 
(color online)
Segregation and spontaneous self-assembly of  $c^-$-rotors in an equimolar binary mixture of non-convex SPPs, cf. \href{https://dl.dropboxusercontent.com/u/101166552/MOVIE_05.mp4}{Supplemental Movie 5}. (a) Snapshot of simulation with $N=2\times 10^{3}$, $\kappa=0.2$, $\alpha = \pi$, $\phi = 0.08$. (b)~Enlarged  view of yellow-shaded area in~(a). (c)
Rotational velocity $\omega = \partial \varphi/ \partial \tau$ indicated by color coding. Insets depict snapshots of clusters composed of $c^-$-crescents rotating clockwise ($\omega <0$) or counter-clockwise ($\omega >0$). (d) A pronounced peak  of the structure factor signals the formation $c^-$-rotors which spin almost ballistically (lower left inset).}
\end{center}
\end{figure}
\par
Similar to the convex case (Figs.~\ref{fig02},~\ref{fig03}), the trajectories  of two or more  colliding crescents depend sensitively on the swimming direction relative to the broken fore-aft symmetry, resulting in distinctly different meso-scale structures [Fig.~\ref{fig04}(a,b)].  For a mixture containing equal numbers of $c^\pm$-SPPs, we again find segregation of the different particle types. Perhaps more importantly, however, the $c^-$-SPPs assemble into clockwise or counter-clockwise spinning rotors   [Fig.~\ref{fig04}(a-c)], characterized by a strongly superdiffusive (almost ballistic) collective rotational motion  with angular MSD~$\langle (\Delta \varphi)^2 \rangle \propto \tau^{\gamma}$ where $\gamma \approx 1.9$  [Fig.~\ref{fig04}(d)]. By contrast, the $c^+$-SPPs show only weak propensity to cluster beyond  pairs or triplets, exhibiting only weakly superdiffusive rotational motion with exponent $\gamma  \approx 1.3$. Test simulations showed that $c^-$-rotors are robust against thermal fluctuations, whereas  the small $c^+$-clusters decay rapidly in the presence of noise. Generally, this basic example illustrates how subtle differences in curvature that break fore-aft symmetry, combined with self-propulsion, can be exploited to self-assemble micro-rotors \cite{2009So,DiLeonardo2010,Schwarz-Linek2013} from linearly moving non-convex objects.

\paragraph*{Discussion \& conclusions.}
In this Letter, we have used large-scale 2D SPP simulations to investigate how particle symmetry affects active collective motion at surfaces and interfaces. SPP simulations have proven useful in the past for understanding generic aspects of collective behavior in  bacterial suspensions~\cite{2012Wensink_PNAS} and other active systems~\cite{2006Peruani,2008BaMa,2012wensinklowen,2012Baskaran,2013Kela}, but such earlier studies focussed almost exclusively on fore-aft symmetric (e.g., spherical or rod-like) particles~\cite{2013Chate_Disks}. Our results show that even subtle violations of fore-aft symmetry can lead to profound changes in the collective dynamics of active colloids or microorganisms, promising new strategies for the self-assembly of active matter. The systematic classification of macroscopic material properties in terms of microscopic constituent symmetries~\cite{1929Pauling,2012Glotzer_Science,2012Graaf_Science,2012Cegeo_EPL} has been very fruitful for the understanding for conventional passive matter. We therefore believe that, with regard to future applications,  it is worthwhile to continue to explore more systematically, both theoretically and experimentally, the interplay between geometric particle symmetries and self-propulsion in active systems.
\par
From a theoretical perspective, simplified shape-based models can provide a useful conceptual link between abstract  Viscek-type vector-based theories~\cite{vicsek1995,farrell2012,2012Pawel_Review} and more accurate microscopic models~\cite{Babu2012}. With regard to microbiological systems, the usefulness of such geometric approaches depends on the collision time-scale of the organisms, which determines whether time-averaged \lq shapes\rq, as shown in Fig.~\ref{fig01}(a,b), are sufficiently accurate approximations to the most relevant interaction effects. However, recent experiments on microbial surface interactions~\cite{2013Kantsler_PNAS}, as well as the results in Figs.~\ref{fig02} and~\ref{fig03},  suggest that shape-induced effects, in conjunction with chemical sensing and hydrodynamic effects, could indeed be a relevant factor in microbial processes such as collective sperm swimming or algal colony formation. Last but not least, in view of the current rapid progress in the fabrication of active colloids~\cite{2013Loewen_PRL} and in the manipulation of microbial shapes~\cite{2009Capeen_EMBO,2013Amir}, we expect that the above results 
can provide useful guidance for the controlled self-assembly of active (e.g., layered or rotating) mesoscale structures  in the near future.
\par
This work was supported by the ERC Advanced Investigator Grant 247333 (R.E.G. and J.D.).

\bibliographystyle{apsrev}

\bibliography{refs}

\end{document}